\documentclass[traditabstract]{aa}
\usepackage{txfonts}
\usepackage{graphicx}
\usepackage{natbib}
\bibpunct{(}{)}{;}{a}{}{,} 
\usepackage{epsfig}

\newcommand{\waspp}{\emph{WASP-8b}}
\newcommand{\wasps}{\emph{WASP-8}}

\newcommand{\kms}{\,km\,s$^{-1}$}
\newcommand{\ms}{\,m\,s$^{-1}$}
\newcommand{\msyr}{\,m\,s$^{-1}$\,yr$^{-1}$}

\newcommand{\vsini}{$v$\,sin\,$i$}

\newcommand{\teff}{T$_{\rm eff}$}
\newcommand{\logg}{log {\it g}}

\newcommand{\mictrb}{\mbox{$\xi_{\rm t}$}}
\newcommand{\halpha}{\mbox{$H_\alpha$}}
\newcommand{\hbeta}{\mbox{$H_\beta$}}
\newcommand{\mactrb}{\mbox{$v_{\rm mac}$}}

\newcommand{\coralie}{\emph{CORALIE }}
\newcommand{\harps}{\emph{HARPS }}
\newcommand{\err}[2] {\ ( $#1\ #2$)}

\begin{document}
\title{WASP-8b: a retrograde transiting planet in a multiple system\thanks{Based on observations made with HARPSspectrograph on the 3.6-m ESO telescope and the EULER Swiss telescope at La Silla Observatory, Chile}\fnmsep\thanks{Radial velocity data are only available in electronic form
at the CDS via anonymous ftp to cdsarc.u-strasbg.fr (130.79.128.5)
or via http://cdsweb.u-strasbg.fr/cgi-bin/qcat?J/A+A/}}
\author{Didier Queloz\inst{1}
\and David Anderson\inst{2}
\and Andrew Collier Cameron\inst{3}
\and Micha\"el Gillon\inst{4} 
\and Leslie Hebb\inst{3}
\and Coel Hellier\inst{2}
\and Pierre Maxted\inst{2}
\and Francesco Pepe\inst{1}
\and Don Pollacco\inst{5}
\and Damien S\'egransan\inst{1}
\and Barry Smalley\inst{2}
\and Amaury H.M.J Triaud\inst{1}
\and St\'ephane Udry \inst{1}
\and Richard West\inst{6}
}

\offprints{Didier.Queloz@unige.ch}

\institute{Observatoire Astronomique de l'Universit\'e de Gen\`eve, Chemin des Maillettes 51, CH-1290 Sauverny, Switzerland
\and Astrophysics Group, Keele University, Staffordshire, ST55BG, UK
\and School of Physics \& Astronomy, University of St Andrews, North Haugh, KY16 9SS, St Andrews, Fife, Scotland, UK
\and Institut d'Astrophysique et de G\'eophysique, Universit\'e de Li\`ege, All\'ee du 6 Ao\^ut, 17, Bat. B5C, Li\`ege 1, Belgium
\and Astrophysics Research Centre, School of Mathematics \& Physics, QueenÕs University, University Road, Belfast, BT71NN, UK
\and Department of Physics and Astronomy, University of Leicester, Leicester, LE17RH, UK
}

\date{Received 12 April 2010 / accepted 21 June 2010}
\authorrunning{Queloz et al.}

\abstract{
We report the discovery of \waspp, a transiting planet of $2.25\pm 0.08$ M$_{\rm Jup}$ on a  strongly inclined eccentric 8.15-day orbit, moving  in a retrograde direction  to the rotation of its late-G host star. Evidence is found that  the star is in a multiple stellar system with two other companions. The dynamical complexity of the system  indicates that it may have experienced 
secular interactions such as  the Kozai mechanism or a formation that differs from  the ``classical" disc-migration theory.}


\maketitle

\section{Introduction}
Transiting planets provide a wealth of information on the structure and formation of planets. The measurement of planet radius  combined with its mass has found a  surprising diversity in the mean densities and in particular ``inflated" hot Jupiters.   Spectroscopic measurement of the Rossiter-McLaughlin effect on the radial velocity during transits indicates that some of these planets may not be aligned with the rotation axes of their stars (see references in   \cite{2010arXiv1001.2010W}). The diversity in the  observed spin-orbit misalignments is somewhat similar to that seen earlier  in period and eccentricity distribution of planets detected by radial velocity surveys (see refs in \cite{2007ARA&A..45..397U} and references therein). The recent sharp rise in the   detections of transiting planets is the outcome of  successful  ground-based wide transit searches surveys among which  WASP \citep{2006PASP..118.1407P}  is the most prolific.   

These discoveries  have stimulated theoretical investigations of alternative formation scenarios to  the  migration theory \citep{1996Natur.380..606L,2003ApJ...589..605W}. These alternative theories
account for  the discoveries of eccentric hot Jupiters on orbits not aligned with the rotation equator of their star \citep{2003ApJ...589..605W, 2009ApJ...696.1230F,2008ApJ...678..498N,2009MNRAS.395.2268B}. 

\section{Observations}\label{sec:obs}
\subsection{The WASP-8 multiple stellar system}
The star \wasps\ (TYC2 7522-505-1) at  $\alpha$(2000): 23h\,59m\,36.07s, $\delta$(2000): $-35^\circ$\,1'\,52.9'', was observed  in 2006 and 2007  by the WASP-south  telescope \citep{2006PASP..118.1407P}. It
is a $V=9.79$ magnitude star with a Tycho (B$-$V) color of  0.73  which is indicative of  a G8 spectral type.  The Infra-red Flux Method (IRFM) \citep{1977MNRAS.180..177B}, using GALEX, TYCHO-2, USNO-B1.0 R-magnitude, and 2MASS broad-band photometry, yields a  distance of 87$\pm7$~pc.

 \wasps\ is identified in the CCDM catalogue  (\object{CCDM 23596-3502})  as  the A component  of a system  of three stars. The B component is a 15th magnitude red star, 4 arcsec south of A, and the third component C is a 10th magnitude star (\object{HIP 118299}, \object{ HD224664}) 142 arcsec north of A. The radial velocity of HD224664 is 4.7 \kms\ and stable over two years  (Mayor priv com.) but differs from the WASP-8 value of  -1.5\kms. The proper motions of the components also differs. It is therefore  unlikely that C and A are physically associated. 

\begin{figure}
\centering                     
\includegraphics[width=9cm]{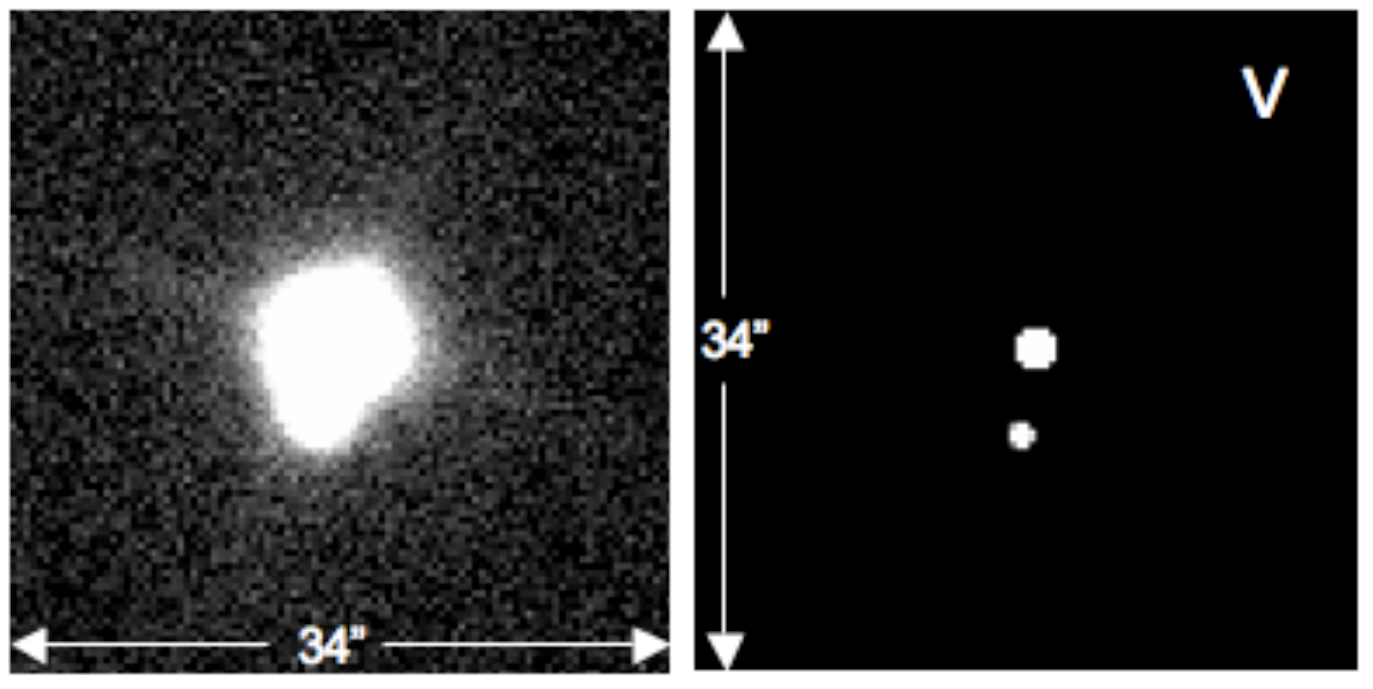}
\caption{Original (left) and deconvolved (right) V-band image from the  Euler telescope of the A and B component of   \wasps\/.}\label{fig:wasp8_ima}
\end{figure}

We measured the photometry and position of \wasps\ and its nearby star  (B component)  with the Euler CCD camera of the 1.2m swiss Euler telescope at La Silla   (see Fig.\,\ref{fig:wasp8_ima}). 
By comparing with nearby stars, we obtained a magnitude difference  $\Delta\mathrm{m}_V=4.7$,  $\Delta\mathrm{m}_I=3.5$.
A separation and a projected angle was measured on the deconvolved images \citep{2007ASPC..366..113G} and we obtained 
$4.83\pm0.01$'' and $\mathrm{PA}=170.7\pm0.1$\degr (only internal errors being considered). Assuming  that \wasps\ and its B component are part of a multiple system, the  color indices would represent those of  an  M star. A similar photometric analysis  of the individual 2MASS archive images  indicates that $\Delta\mathrm{m}_J=2.7$,  $\Delta\mathrm{m}_H=2.2$, and  $\Delta\mathrm{m}_K=2.1$, which are also indicative of  an M star.  The value mentioned in the Washington Visual Double Star Catalog  measured 70 years ago indicates  4.0" and PA=170\degr \citep{2001AJ....122.3466M}. This suggests little, if any, relative motion of the two stars over the 70-year time span between these observations. When compared with the proper
motion in right ascension of \wasps\/, about 100 mas/yr \citep{2004AAS...205.4815Z}, this indicates a common proper motion pair. Given the distance of \wasps\, the  sky-projected separation of the pair is about 440\,AU. Using   available differential photometry, we estimate the temperature of the  B component to be about  $3700$\,K.

\subsection{Photometric and radial-velocity observations}

\begin{figure}
\centering                     
\includegraphics[width=9cm]{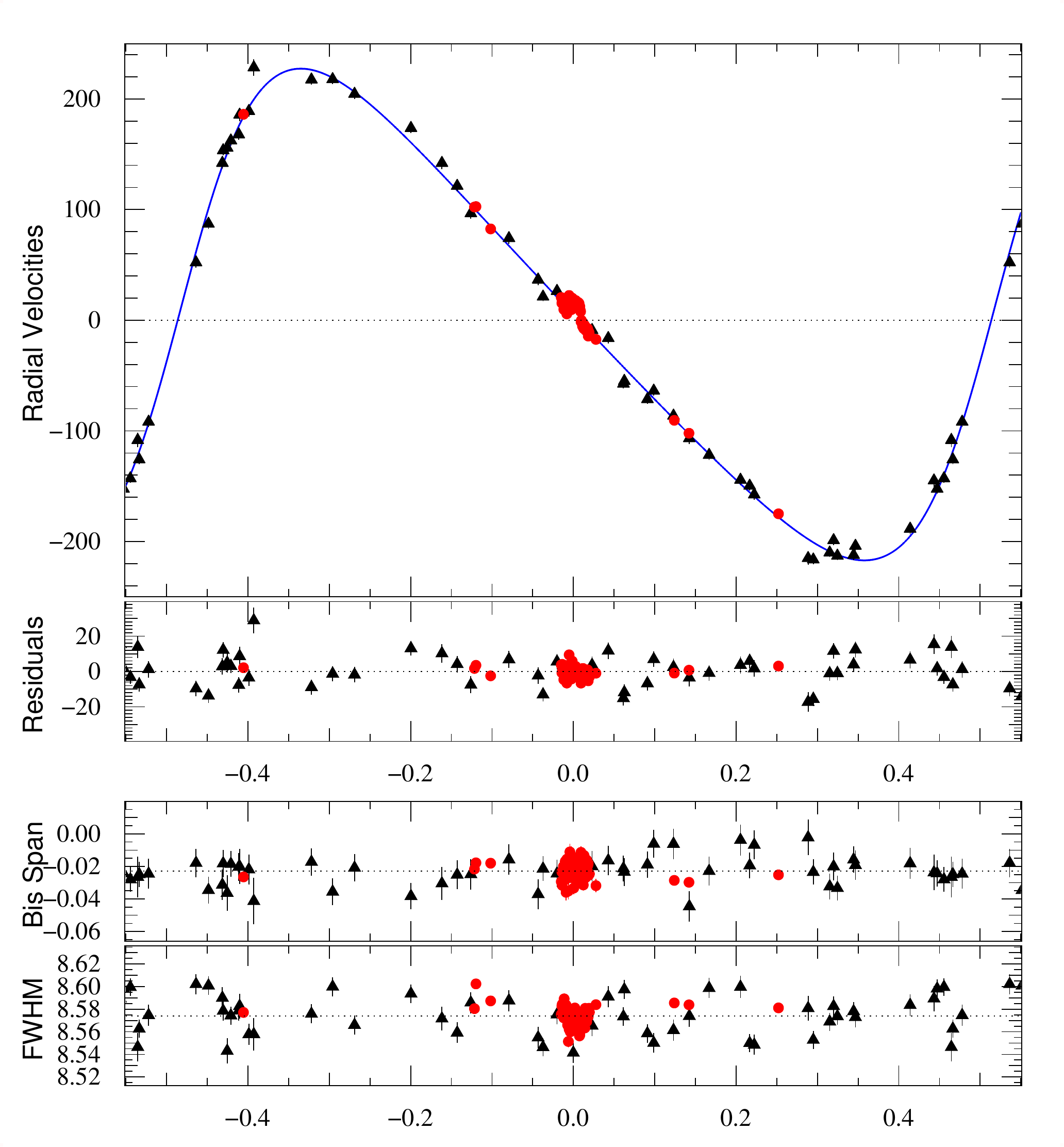}
\caption{{\bf Top}: Overall Keplerian fit to the RV data for Wasp-8b.
Black triangles indicates \textit{Coralie} data and red dots the  \textit{Harps} data. The long-term drift was removed  to plot the velocity in phase (the zero  is set  at the time of the transit).
{\bf Bottom}: Bissector span and FWHM (in \kms unit) plotted  with the phase of the orbit. The \harps radial velocity data were shifted   to correct from the $\gamma$ velocity  difference with  \coralie. \label{fig:wasp8full}}
\end{figure}

\begin{figure}
\centering                     
\includegraphics[width=9cm]{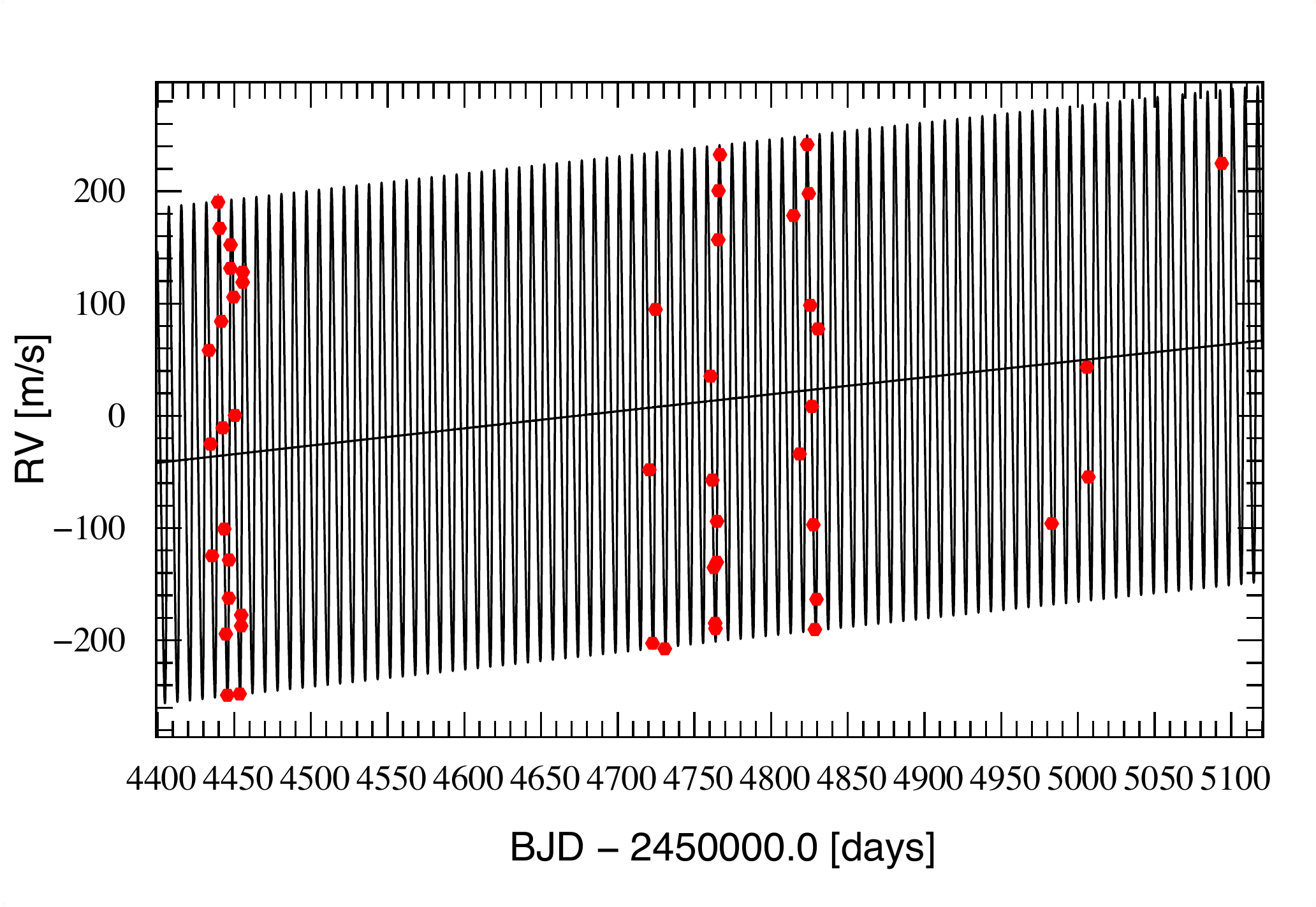}
\caption{\coralie radial velocity measurements  (red dots) of \wasps\  superimposed on the best-fit solution (solid line). }\label{fig:wasp8drift}
\end{figure}

\wasps\  was recorded simultaneously by two cameras of the WASP-south telescope  during two seasons (2006 and 2007). Altogether 11\,224 independent photometric points were recorded with a typical sampling of 8 minutes. Transit events were detected in  data from the first observation season. This triggered  radial velocity  follow-up observations of \wasps\   in November 2007 with the \textit{Coralie} spectrograph mounted at  the Swiss Euler telescope \citep{1996A&AS..119..373B,2000A&A...354...99Q,2002A&A...388..632P}.
With a combined analysis of the radial-velocity data and the photometry including additional  WASP data from the 2007 season, a transit period of 8.15\,days was  found. No  changes to the spectroscopic profile were detected, ruling out a blended eclipsing binary or starspots as the cause of the radial-velocity variation \citep{2001A&A...379..279Q} (see bottom diagram in Fig~\ref{fig:wasp8full}). In the  next  season, observations with \coralie were continued,  revealing an additional drift in the $\gamma$ velocity of the system (Fig.~\ref{fig:wasp8drift}). No second-order curvature term was detected.

\begin{figure}
\centering                     
\includegraphics[width=9cm]{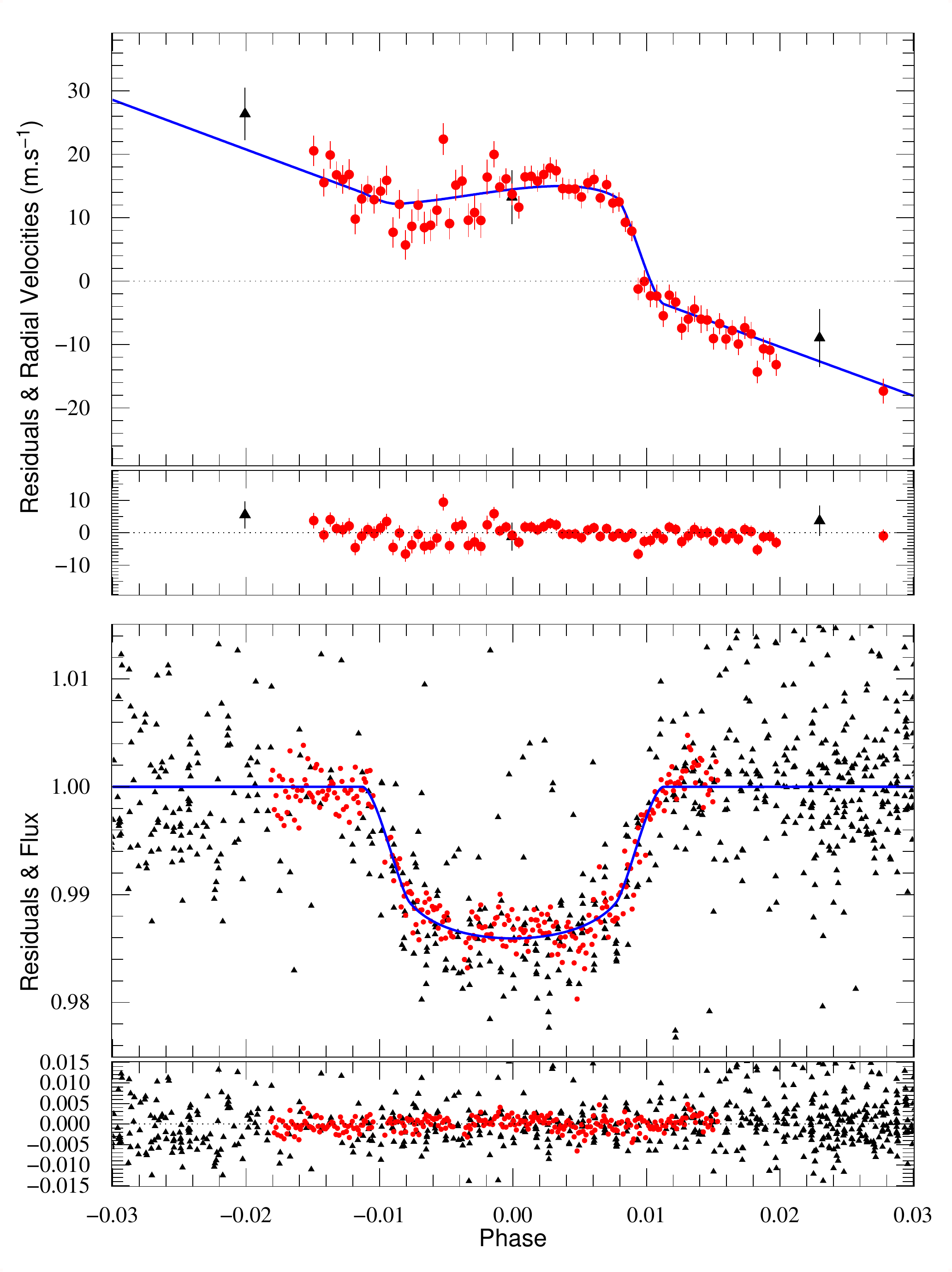}
\caption{{\bf{Top}}: Radial velocity measurement phased with the transit (mid-transit is at 0). Black triangles are \coralie data and red dots \harps data.
{\bf{Bottom}}: Normalized transit photometry measurement of \wasps. Black triangles indicates SuperWASP data and red dots the R-band Euler photometry data. The best-fit model is superimposed in blue.}\label{fig:wasp8trans}
\end{figure}

On 25 August 2008,  following up on the confirmation of the planet, a  complete and densely  sampled transit event was recorded  in R band with the Euler  telescope to improve the  determination of  the transit parameters (Fig.~\ref{fig:wasp8trans}).   On 4 October 2008,  a spectroscopic transit was measured with the \harps spectrograph installed on the 3.6m telescope at La Silla. During the sequence,  75 spectra (44  in the transit) were measured with an exposure time of 300\,s, corresponding to a  typical signal-to-noise ratio per pixel of 50. The radial velocity measurement  from these spectra shows  an obvious  Rossiter-McLaughlin effect  with a shape suggesting a non-coplanar orbit (Fig.~\ref{fig:wasp8trans}).  In addition, four spectra were measured on the same night before and after the transit  to help us to determine the rate of change in the radial velocity  outside the transit.  Three  measurements  were obtained later at other phases of the system to improve the matching and zero point correction between \coralie and \harps data. During the measurement of the transit sequence,  a significant change in telescope focus happened at JD 54744.592, improving the  flux entering the fiber by a factor of 2.

\section{Determination of system parameters}\label{sec:analysis}

\subsection{Spectral analysis}

The individual \harps spectra  were coadded and used for a detailed spectroscopic analysis of WASP-8. The results are displayed in Table~\ref{wasp8-params}. As in previous WASP-papers  \citep{2007MNRAS.375..951C}, the analysis was performed using the {\sc uclsyn} spectral synthesis package  and {\sc atlas9} models without convective overshooting  with  \halpha\ and \hbeta\ Na {\sc i} D and Mg {\sc i} b lines  as diagnostics for \teff\ and  (\logg). The abundances and the microturbulence   were determined in a similarly way to the work of  \cite{2009A&A...496..259G} and  used as additional \teff\ and \logg\ diagnostics \citep{2005MSAIS...8..130S}. 

The Li {\sc i} 6708\AA\ line is detected in the spectra indicating an abundance of {log A(Li/H) + 12} = 1.5
$\pm$ 0.1, which implies an age of  3-5Gyr according to \cite{2005A&A...442..615S}.  However, \cite{2009Natur.462..189I}  noted that stars with planets have lower lithium abundances than normal solar-type stars, so the lithium abundance may not be a good age indicator for them.

The rotational broadening \vsini\  was measured by fitting the
observed HARPS profiles of several unblended Fe~{\sc i} lines. A typical value of
macroturbulence \mactrb$ = 2$ \kms\ was adopted and an
instrumental profile determined from telluric absorption
lines. We found that  \vsini\ = 2.0 $\pm$
0.6 \kms, which is  typical of  a G dwarf of intermediate age.

\begin{table}
\caption{Stellar parameters of \wasps\ derived from spectroscopic analysis. The quoted error
estimates include those given by the uncertainties in \teff, \logg, and \mictrb,
as well as  atomic data uncertainties.}
\begin{tabular}{ll|cc}
\hline\hline                 
\noalign{\smallskip}
\teff      & 5600 $\pm$ 80 K&{[Na/H]}   &  +0.22 $\pm$ 0.07 \\
\logg      & 4.5 $\pm$ 0.1 &{[Mg/H]}   &  +0.21 $\pm$ 0.04 \\
\mictrb    & 1.1 $\pm$ 0.1 \kms& {[Si/H]}   &  +0.29 $\pm$ 0.09 \\
\vsini     & 2.0 $\pm$ 0.6 \kms & {[Ca/H]}   &  +0.24 $\pm$ 0.12\\
{[Fe/H]}   &  +0.17 $\pm$ 0.07  &  {[Sc/H]}   &  +0.23 $\pm$ 0.05  \\
log A(Li/H)+12  & 1.5 $\pm$ 0.1&{[Ti/H]}   &  +0.24 $\pm$ 0.08  \\
 \
&&{[V/H]}    &  +0.30 $\pm$ 0.08 \\
dist&87$\pm7$ pc&{[Cr/H]}   &  +0.17 $\pm$ 0.09 \\
age&3-5\,Gyr&{[Co/H]}   &  +0.29 $\pm$ 0.07 \\
&&{[Ni/H]}   &  +0.23 $\pm$ 0.07 \\
\noalign{\smallskip}
\hline\hline                 
\end{tabular}
\label{wasp8-params}
\end{table}

\subsection{Analysis  of the planetary system}

\begin{figure}
\centering                     
\includegraphics[width=9cm]{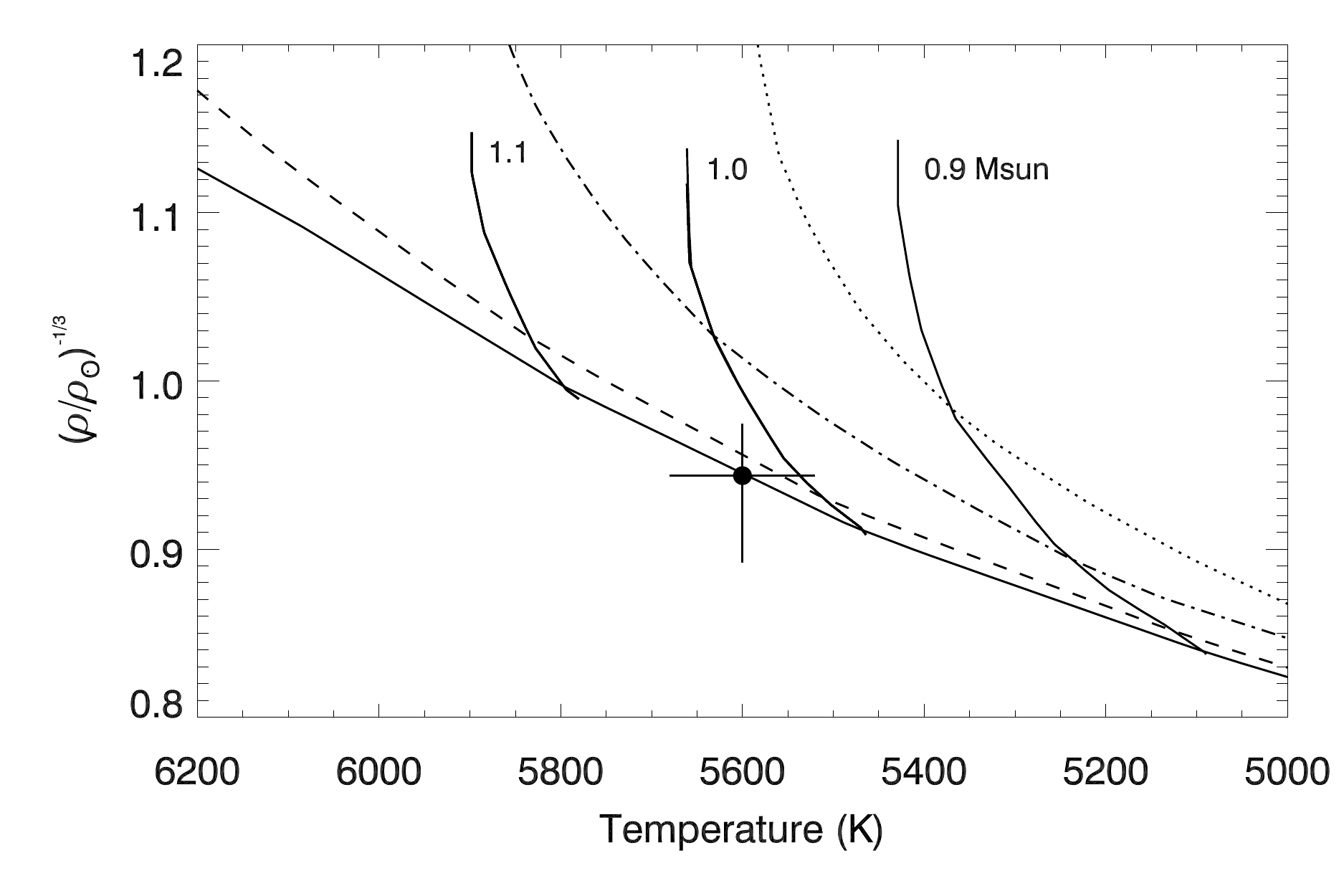}
\caption{Comparison of the best-fitting stellar parameters from the transit profile and spectroscopic analysis with evolutionary models interpolated at [Fe/H]=0.17.  The isochrones are 100 Myr, 1 Gyr, 5 Gyr and 10 Gyr.  The evolutionary tracks are indicated for  0.9, 1.0 and 1.1 Msun.}\label{fig:evol}
\end{figure}

This whole data set   was found to detect without doubt  a planet transiting the star \wasps\/. 
We analyzed  together the photometric (WASP and Euler data)  and the radial velocity data, including the spectroscopic transit sequence in this context. Our model  was based on the  transit modeling by \cite {2002ApJ...580L.171M}  and the radial velocity description by \cite{2006ApJ...650..408G}. The  best-fit model parameters and their error bars were computed  using a MCMC convergence scheme that solves all parameters together.  For details of the code  and fitting techniques, we refer to \cite{2009A&A...506..377T,2007MNRAS.380.1230C}. To obtain a coherent solution, we determined the mass of the star by comparing the spectroscopically-determined effective temperature and the stellar density outcome of the MCMC adjustment, with evolutionary tracks and isochrones of the observed metallicity from  the stellar evolution model of \cite{2000A&AS..141..371G}. We converged iteratively on a stellar mass of $1.04 $(+0.02 -0.09)$ \,M_{\odot}$   and an age younger than 6\,Gyr (see on Fig.~\ref{fig:evol}).

The free  parameters of our model were the depth of transit $D$, the width of transit $W$, the impact parameter $b$, the period $P$, the epoch of transit centre $T_0$, the RV semi-amplitude $K$,  $e\,cos\,\omega$ \& $e\,sin\,\omega$ ($e$ being the eccentricity and $\omega$ the angle of the periastron), and $V\,sin\,I\,cos \,\beta$ \& $V\,sin\,I\,sin\,\beta$, with $V\,sin\,I$ being the projection of the stellar equatorial rotation, and $\beta$ the projection of the angle between the stellar spin axis and the planetary orbit axis. In addition, we employed free normalization factors for each lightcurve (WASP and Euler) and  each set of radial velocity ($\gamma_\textit{\tiny H}$ for \harps and  $\gamma_\textit{\tiny C}$ for \coralie), which enabled variations to be made in  instrumental zero points. From these  parameters,  {\sl physical} parameters were derived  to characterise the planetary system.  The  best-fit set of parameters that minimize the  $\chi^2_r$ (reduced $\chi^2$  is 0.86)  are listed in Table~\ref{tab:params} as well as  their related computed physical parameters. With this best-fit solution  one computes for the \coralie data $\chi^2=204$ with 48 measurements, and for  HARPS data $\chi^2=188$ with 82 measurements which implies that additional jittering is present that is 
not accounted for by the fitted model. Since the main  deviation is  related to  the \coralie data,  the uncertainties in the orbital solutions are most likely underestimated. However, the error bars in the Rossiter parameters are  driven mostly by the HARPS on-transit data, and one can assume that they are almost correct.

\begin{table}
\caption{Fitted and physical parameters of the \wasps\ planetary system. The error bars are calculated  at 68\,\% of the statistical distributions.
}\label{tab:params}
\begin{tabular}{lrl}
\hline
\hline\\[-3mm]
$D$                                                                      & 0.01276           &       \err{+ 0.00033}{- 0.00030}    \\
$W$ (days)                                                           & 0.1832             &     \err{+ 0.0030}{- 0.0024 }   \\
$b$ $(R_{\star})$                                                   & 0.604               &       \err{+ 0.043}{- 0.040}    \\
$P$ (days)                                                            & 8.158715         &     \err{+0.000016}{-0.000015}   \\
$T_0$ (BJD-2450000)                                            & 4679.33394      &\err{+ 0.00050}{- 0.00043}     \\
$K$ (\ms)                                                             & 222.23             &\err{+ 0.84}{- 0.60}  \\
d$\gamma$/dt (\msyr)                                           & 58.1                 &\err{+1.2}{-1.3} \\
$e\,cos\,\omega$                                                  & 0.02307           &\err{+ 0.0010}{- 0.0010}  \\
$e\,sin\,\omega$                                                   & $-0.3092$        & \err{+ 0.0024}{- 0.0029}  \\
$V\,sin\,I\,cos\,\beta$                                            & $-0.873$          &\err{+ 0.059}{- 0.064}     \\
$\gamma_\textit{\tiny C}$\tablefootmark{a}   (\ms)    & $-1565.76$      &\err{+  0.16 }{-0.21}\\
$\gamma\textit{\tiny H}$\tablefootmark{a}   (\ms)      & $-1548.10$      &\err{+  0.60 }{-0.13}\\
$V\,sin\,I\,sin\,\beta$                                             & 1.59                &\err{+ 0.08}{- 0.09}    \\
$R_p / R_{\star}$                                                   &  0.1130           &\err{+ 0.0015}{- 0.0013}   \\
$R_{\star} / a$                                                       &  0.0549           &\err{+ 0.0024 }{- 0.0024}   \\
$\rho_{\star}$   $(\rho_{\odot})$                                &  1.22              &\err{+ 0.17 }{- 0.15}          \\
$R_{\star}$  $(R_{\odot})$                                       & 0.945              &\err{+0.051}{-0.036}                 \\
$M_{\star} $  $(M_{\odot})$                                     &  1.030             &\err{+ 0.054 }{- 0.061}          \\
$V\,sin\,I$  $(km.s^{-1})$                                       &1.59                 &\err{+0.08}{-0.09}\\
$R_p / a$                                                             &  0.00620          &\err{+ 0.00036 }{- 0.00033}   \\
$R_p$  $(R_J)$                                                    &  1.038             &\err{+ 0.007}{- 0.047}              \\
$M_p$ $(M_J)$                                                    & 2.244              &\err{+ 0.079}{- 0.093}                     \\
$a$ $(AU)$                                                          &  0.0801            &\err{+ 0.0014}{- 0.0016}      \\
$i$   $(^{\circ})$                                                    &  88.55              &\err{+ 0.15}{- 0.17}                        \\
$e$                                                                     &   0.3100            &\err{+ 0.0029}{- 0.0024}                 \\
$\omega$  $(^{\circ})$                                           & $-85.73$          &\err{+ 0.17}{- 0.18}                              \\
$\beta$ $(^{\circ})$                                               & 123.3               &\err{+4.4}{-3.4}                               \\
\hline
\hline

\end{tabular}
\tablefoottext{a}{computed at JDB=2454691.15781}
\end{table}

Our best-fit solution corresponds to a  giant planet  with an eccentric ($e=0.3$) 8.16-day  orbit and an additional  long-term radial-velocity drift of 58\msyr. The planet is dense with $2.25\,M_j$ and a radius of $1.04\,R_j$, in contrast to the substantial  fraction of  `inflated"  hot Jupiters.  Surprisingly, the projected angle between the orbital and stellar spin axes is found to be $\beta = 123.3^\circ$,  indicative of a  retrograde orbit.   We note that $V\,sin\,I=1.59$\kms\ is in accordance with  the  line rotation broadening    \vsini\  (in Table~\ref{wasp8-params}) derived by the spectral analysis. 

We checked whether the partial defocusing of  HARPS during the transit spectroscopic sequence had any effect on our result.  We  divided  the series  into two subsets and considered for each of them an  independent offset ($\gamma$). We obtain a solution with a marginal improvement in the    $\chi^2$. By comparing  the solution obtained from  these two sets with that for the complete set, the angle $\beta$ was changed by $1.5\,\sigma$. The defocusing problem  does not affect the results of this paper.

\section{Discussion}
The detection of a hot Jupiter on an eccentric orbit that is misaligned with the stellar rotation axis and moving in a  retrograde direction raises many questions about the origins of this system. Although the answer is beyond the scope of this paper,  the visual faint companion  and  the drifting $\gamma$ velocity of the system are key components of the puzzle. From the observed separation between the A and B components, one can derive a most likely  orbital semi-major axis ($a=1.35\rho\approx600$\,AU)   \citep{1991A&A...248..485D}. The observed radial-velocity drift is therefore unlikely to be related to the  B component  of the binary  ($\dot{\gamma}<GMa^{-2}<1$\msyr), suggesting that these is an additional closer companion of both unknown mass and period. 
The lack of curvature indicates that the  companion is more massive than the transiting planet.
This intermediate body is very likely to play a significant dynamical role in the system.

Apart  from the complex dynamics of the whole system, the planet  \waspp\ is a   "standard" hot Jupiter. It orbits a metal-rich star, which accounts for the observed increase in the incidences of hot Jupiters  with  the metallicity of the host star \citep{2007ARA&A..45..397U}. The period of \waspp\  is longer than the 3-4 days typical value, but considering the eccentricity of its orbit, its periastron distance  is  typical of hot Jupiters. 

The orbit misalignment of the planet  with the stellar rotation axis of \wasps\   is measured with the $\beta$ parameter. The true angle between the axes of the stellar and planetary orbits is  usually called $\psi$ and is statistically related to $\beta$ through $\sin I$ (unknown) and the orbital inclination  ($i$) (see \cite{2009ApJ...696.1230F} for details). When $\beta$ deviates significantly from zero, this provides us with a lower limit to the $\psi$. When $\beta$ is beyond 90$^\circ$, the orbital spin has the  opposite direction  to the stellar rotation provided that  the orbit does not transit the star between its pole and its limb. According to Eq.\,9 from \cite{2009ApJ...696.1230F}, this condition is met when  $I>3.6$ degree. By combining   $V\,sin\,I,$ with  the estimated age of the star, one can exclude such a small $I$ angle. Interpreted with the large $\beta$ value, we can conclude that a true retrograde orbit is the most likely scenario  for \waspp. 

The origin of the unusual shape and orientation of the orbit of \waspp\ is possibly related to   the Kozai mechanism  \citep{1962AJ.....67..591K, 2003ApJ...589..605W} or the outcome of a  violent dynamical interaction history. The evidence of two other bodies and a possible series of  secular effects \citep{2008ApJ...683.1063T} make the \wasps\ system unique and interesting for additional dynamical studies and  a test case for formation scenarios of hot Jupiters  that constitute an alternative to the disc-migration mechanism.


\bibliographystyle{aa}
\bibliography{bibtex}

\end{document}